\documentclass[useAMS,usenatbib]{mn2e}

\usepackage{graphicx}
\usepackage{txfonts}
\usepackage{natbib}
\newcommand\aj{AJ}
\newcommand\apj{ApJ}

\newcommand\aap{A\&A}
\newcommand\mnras{MNRAS}
\newcommand\apjl{ApJ}
\newcommand\pasp{PASP}
\newcommand\nat{Nature}

\title[]{The {\it Hubble Space Telescope} UV Legacy Survey of Galactic Globular Clusters.
II. The seven stellar populations of NGC\,7089
(M\,2)\footnote{Based on observations with the NASA/ESA {\it Hubble Space Telescope}, obtained at the Space Telescope Science Institute, which is operated by AURA, Inc., under NASA contract NAS 5-26555.}
\author[A.\,P.\, Milone et al.\,]
{A.\,P.\,Milone$^{1}$,
 A.\,F.\,Marino$^{1}$,
 G.\,Piotto$^{2,3}$,
 L.\,R.\,Bedin$^{3}$,
 J.\,Anderson$^{4}$,
 A.\,Renzini$^{3}$,
 \newauthor
 I.\,R.\,King$^{5}$,
 A.\,Bellini$^{4}$,
 T.\,M.\,Brown$^{4}$,
 S.\,Cassisi$^{6}$,
 F.\,D'Antona$^{7}$,
 H.\,Jerjen$^{1}$,
 D.\,Nardiello$^{1,2}$,
\newauthor
 M.\,Salaris$^{8}$,
 R.\,P.\,van der Marel$^{4}$,
 E.\,Vesperini$^{9}$,
 D.\,Yong$^{1}$,
 A.\,Aparicio$^{10,11}$,
\newauthor
 A.\,Sarajedini$^{12}$,
 M.\,Zoccali$^{13,14}$\\
$^{1}$Research School of Astronomy \& Astrophysics, Australian National University, Mt Stromlo Observatory, via Cotter Rd, Weston, ACT 2611, Australia \\
$^{2}$Dipartimento di Fisica e Astronomia ``Galileo Galilei'', Univ. di Padova, Vicolo dell'Osservatorio 3, Padova, IT-35122\\
$^{3}$Istituto Nazionale di Astrofisica - Osservatorio Astronomico di Padova, Vicolo dell'Osservatorio 5, Padova, IT-35122\\
$^{4}$Space Telescope Science Institute, 3800 San Martin Drive, Baltimore,  MD 21218, USA\\
$^{5}$Department of Astronomy, University of Washington, Box 351580, Seattle, WA 98195-1580\\
$^{6}$Istituto Nazionale di Astrofisica - Osservatorio Astronomico di Teramo, Via Mentore  Maggini s.n.c., I-64100 Teramo, Italy\\
$^{7}$Istituto Nazionale di Astrofisica - Osservatorio Astronomico di Roma, Via Frascati 33, I-00040 Monteporzio Catone, Roma, Italy\\
$^{8}$Astrophysics Research Institute, Liverpool John Moores University, Liverpool Science Park, IC2 Building, 146 Brownlow Hill, Liverpool L3 5RF, UK\\
$^{9}$Department of Astronomy, Indiana University, Bloomington, IN 47405, USA\\
$^{10}$Instituto de Astrof\`isica de Canarias, E-38200 La Laguna, Tenerife, Canary Islands, Spain\\
$^{11}$Department of Astrophysics, University of La Laguna, E-38200 La Laguna, Tenerife, Canary Islands, Spain\\
$^{12}$Department of Astronomy, University of Florida, 211 Bryant Space Science Center, Gainesville, FL 32611, USA\\
$^{13}$Universidad Cat\`olica de Chile, Departamento de Astronom\`ia y Astrof\`isica, Casilla 306, Santiago 22, Chile\\
 $^{14}$Millennium Institute of Astrophysics, Av Vicuna Mackenna 4860, Macul, Santiago, Chile
   }}

\begin{document}
\date{Draft Version Nov, 7, 2014}

\pagerange{\pageref{firstpage}--\pageref{lastpage}} \pubyear{2013}

\maketitle
\label{firstpage}

\begin{abstract}  
We present high-precision multi-band photometry for the globular cluster (GC) M\,2. We combine the analysis of the photometric data
 obtained from the {\it Hubble Space Telescope} UV
Legacy Survey of Galactic GCs GO-13297, 
 with chemical abundances by Yong et al.\,(2014), and compare the photometry
with models in order to  analyze the multiple stellar sequences we identified in the color-magnitude diagram (CMD). 
We find three main stellar components, composed of metal-poor, metal-intermediate, and metal-rich stars  (hereafter 
referred to as population A, B, and C, respectively).
 The components A and B include
stars with different $s$-process element abundances.
 They host six sub-populations with different light-element abundances, and exhibit an internal variation in helium up to $\Delta$Y$\sim$0.07 dex. 
In contrast with M\,22, another cluster  characterized by the presence of populations with different metallicities, M\,2 contains a third stellar component, C, which shows neither 
evidence for sub-populations nor an internal spread in light-elements. Population C
does not exhibit the typical
photometric signatures that are associated with abundance variations of light elements produced by hydrogen burning at hot temperatures.
We compare M\,2 with other GCs with intrinsic heavy-element 
variations and conclude that M\,2 resembles
M\,22, but it includes an additional stellar component that makes it more similar to the central region of the Sagittarius galaxy, which hosts 
a GC (M54) and the nucleus of the Sagittarius galaxy itself.
\end{abstract}


\section{Introduction}\label{sec:intro}
The {\it Hubble Space Telescope} ({\it HST\/}) 
``Legacy Survey of Galactic Globular Clusters:
Shedding UV Light on Their Populations and Formation'' is designed 
to image 47 Galactic Globular Clusters (GCs) through the filters F275W, F336W and F438W of the ultraviolet and visual channel (UVIS) of the Wide Field Camera 3 (WFC3) on board of {\it HST\/} (GO-13297, PI.\,G.\,Piotto, see Piotto et al.\,2014 --- paper\,I hereafter --- for details). GO-13297
will complement the F606W and F814W database from the Advanced Camera for Survey (ACS) GC Treasury program (GO-10775, PI. A.\,Sarajedini, see Sarajedini et al.\,2007) to provide homogeneous and accurate photometry of GCs in five bands, from  $\sim$275 to $\sim$814 nm. The main goal of this project is to detect and characterize multiple stellar populations in GCs. Indeed, these
five filters have shown a high sensitivity to abundance variations including light-element and helium (see Milone et al.\,2012a, 2013 and Paper\,I for details). 

In the present paper we 
focus on NGC\,7089 (M\,2), which is one of the few GCs that
exhibits a split sub giant branch (SGB) in the visual color-magnitude diagram (CMD, see the \,$m_{\rm F606W}$ vs.\,$m_{\rm F606W}-m_{\rm F814W}$ CMD in
Piotto et al.\,2012). 
The $U$ vs.\,$U-V$ CMD of M\,2 shows a poorly-populated red-giant branch (RGB) on the redward side of the main RGB (Grundahl et al.\,1999; Lardo et al.\,2012). Spectroscopy has revealed that stars in the two RGBs have different abundances in terms of their overall metallicity (Yong et al.\,2014, hereafter Y14) and in terms of their $s$-process elements, with the red-RGB being also $s$-rich (Lardo et al.\,2013; Y14).
More specifically, Y14 have shown that the metallicity distribution of M\,2 stars has three peaks, around [Fe/H]$\sim -$1.7, $-$1.5, and $-$1.0. 
Therefore M\,2 also is one of the very few Galactic GCs with a   large intrinsic difference in iron abundance  ($\Delta$[Fe/H]$\gtrsim$0.1~dex), where Supernovae (SNe) have likely played a major role in the internal chemical-enrichment history.\footnote{ Small star-to-star iron variations, at the level of $\lesssim$0.05 dex are likely present in most GCs and can been detected from high-precision spectroscopic measurement (Yong et al.\,2013).}
The other presently known members of this short list are:
$\omega$\,Centauri, M\,22, Terzan\,5, M\,54, and NGC\,5824 (e.g.\,Norris \& Da Costa\,1995; Johnson \& Pilachowski\,2010; Marino et al.\,2009, 2011a,b; Da Costa et al.\,2009, 2014; Ferraro et al.\,2009;  Carretta et al.\,2010a,b). NGC\,1851 is an intriguing candidate as it exhibits two main stellar populations with distinct $s$-element abundance and a difference in [Fe/H] of $\sim$0.05 dex (Yong et al.\,2008; Carretta et al.\,2010c; Gratton et al.\,2012; Marino et al.\,2014).
 Following Marino et al.\,(2012), we will refer to these non-mono-metallicity GCs as `anomalous'. 
This paper describes the stellar-population properties of one of these anomalous GCs: M\,2. 
We combine high-precision multi-wavelength photometry from GO-13297 and GO-10775
with information from high-resolution spectroscopy to investigate multiple stellar populations in M\,2 and understand their formation and evolution.  The paper is organized as follows: in Sect.~\ref{sec:data} we describe the data and the reduction procedures. The $m_{\rm F275W}$ vs.\,$m_{\rm F275W}-m_{\rm F814W}$ CMD of M\,2 is discussed in Sect.~3, while in Sect.~4 and Sect.~5 we identify the stellar populations along the RGB and the MS, respectively, and infer their abundance of iron and light elements. We estimate the age and the helium content of multiple stellar populations in Sect.~6 and compare M\,2 with other anomalous GCs in Sect.~7. Summary and conclusions are given in Sect.~8.

\section{Data and data analysis}
\label{sec:data}
In order to investigate multiple stellar populations in M\,2 we have 
used images taken with  ACS/WFC and WFC3/UVIS on board the {\it HST\/}. 
The ACS/WFC dataset consists of 5$\times$340$s$ long exposures in F606W and F814W plus one 20$s$ short exposure taken through each of the same filters. These images were taken on April, 16, 2006 as part of the ACS {\it HST\/} Treasury Survey of GCs (GO\,10775, PI.\,A.\,Sarajedini). 
The astro-photometric catalog of stars in the ACS/WFC field used in this paper has been published by Anderson et al.\,(2008).

The WFC3/UVIS dataset includes 6$\times \sim$700$s$ exposures in F275W, 6$\times \sim$300$s$ in F336W, and 3$\times\sim $60$s$ in F438W. These images were taken on August, 14, August, 29, and October, 18, 2013 as part of GO-13297.
 These data were corrected for charge transfer deficiencies by using an algorithm that was developed specifically for UVIS and is based on the method and the software of Anderson \& Bedin\,(2010).
The reduction has been performed with img2xym\_WFC3, software developed by Bellini et al.\,(2010) mostly based on img2xym\_WFI (Anderson et al.\,2006). Photometry and astrometry have been independently carried out for each exposure by using a set of spatially-variable empirical PSFs. We have corrected the stellar positions for geometrical-distortion by using the solution from Bellini et al.\,(2009, 2011) and have calibrated the photometry into the VEGA-mag flight system as in Bedin et al.\,(2005). 
Stellar proper motions have been obtained as in Anderson \& King\,(2003) by comparing the average stellar positions measured from GO-10775 and from GO-13297 data.  Our data cover a temporal baseline of 7.5 yr. Since the cluster has been centered in both the ACS/WFC and WFC3/UVIS fields of view, five-band photometry and proper motions are available for stars in a $\sim$2.7$\times$2.7 arcmin region around the center of M\,2. 

As the photometric separation among sequences populated by different stellar populations is typically small, the study of multiple populations along the CMD of any GC requires very accurate photometry. 
Because of this, we have selected the best-measured stars in our sample, by following the same selection criteria as described in Milone et al.\,(2009), which are based on several diagnostics, including the amount of scattered light from neighboring stars, PSF-fit residuals, and rms scatter in position measurements.
Photometry has been corrected for differential reddening by using the method 
explained in detail in Milone et al.\,(2012a). 
 Briefly, we identified for each star the 55 nearest well-measured main-sequence (MS) stars and determined the color distance from the MS ridge line along the reddening direction. The median color distance of the 55 neighbors 
 has been assumed as the best estimate of differential reddening for  each target star,  and has been applied as a correction to its color.  The reddening in the direction of NGC\,7089 is E(B$-$V)$\sim$0.06 mag, (Harris\,1996 updated as in 2010, see also Sect.~6). We have found that reddening variations in the analyzed field of view are, on average, $\Delta$E(B$-$V)$\sim$0.007 mag and never exceed 0.021 mag.

\section{The F275W vs. F275W$-$F814W CMD}
\label{sec:CMD}
\begin{centering}
\begin{figure*}
 \includegraphics[width=14.5cm]{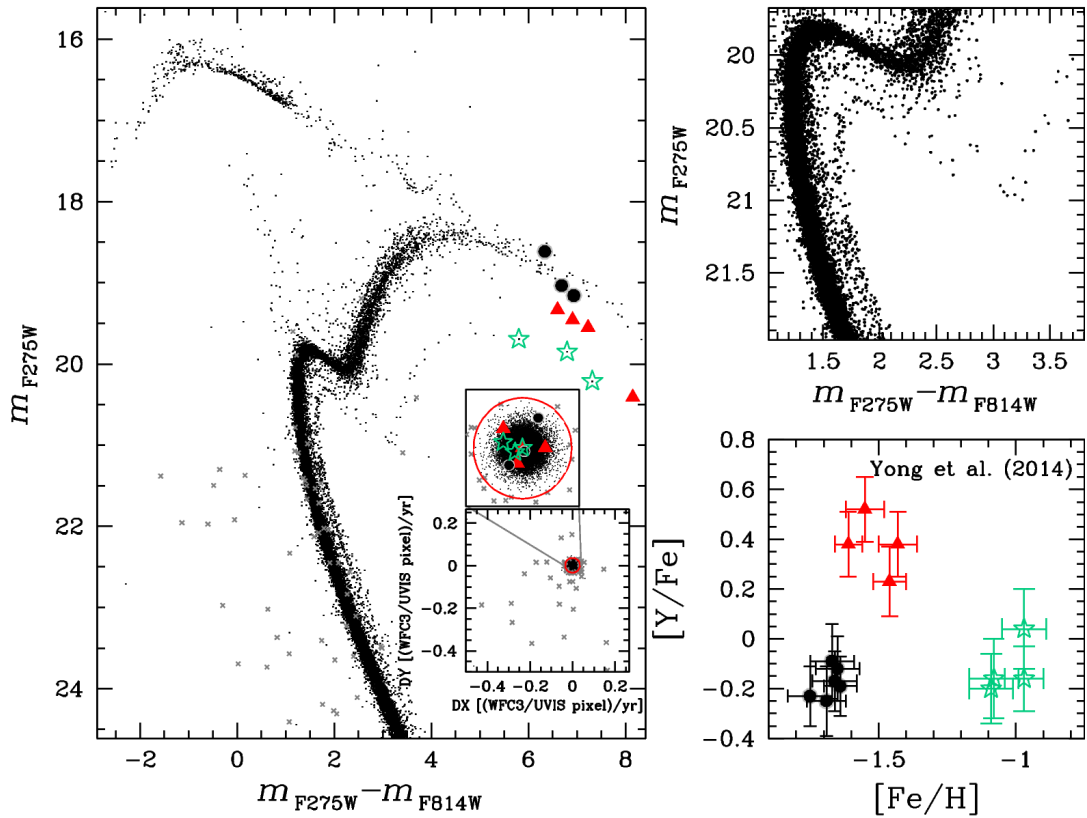}
 \caption{\textit{Left panel}: $m_{\rm F275W}$ vs.\,$m_{\rm F275W}-m_{\rm F814W}$ CMD of stars  in the WFC3/UVIS field of view centered on M\,2. Cluster members and field stars, selected on the basis of their proper motions,  are represented with black dots, and gray crosses, respectively. 
 The vector-point diagram of proper motions is plotted in the inset and the stars within the red circle are considered as cluster members.
A zoom-in of the same CMD around the SGB is plotted in the upper-right panel. \textit{Lower-right panel}: [Y/Fe] vs.\,[Fe/H] for RGB stars from Y14. Metal-poor, metal-intermediate, and metal-rich stars are represented as
black circles, red triangles, and aqua stars, respectively. 
In the \textit{left panel} we used the same  symbols to highlight the Y14 stars we could cross-identify in our photometric catalog. }
 \label{fig:NGC7089cmd1003}
\end{figure*}
\end{centering}

The left panel of Fig.~\ref{fig:NGC7089cmd1003} shows the complete 
$m_{\rm F275W}$ vs.\,$m_{\rm F275W}-m_{\rm F814W}$ CMD of M\,2 members  (black dots), and field stars (gray crosses), selected on the basis of their proper motions.
 The vector-point diagram (VPD) of proper motions in WFC3/UVIS pixel units per year is plotted in the inset and includes all the stars plotted in the CMD. Since we have used M\,2 members as reference stars to calculate proper motions, the bulk of stars around the origin of the VPD is mostly made of clusters members while field stars have larger proper motions. We have thus drawn a red circle to separate the probable cluster members from field stars. In this paper we analyze cluster members only. To identify them, we have chosen a radius of 0.03 pixel/yr that corresponds to five times the average proper-motion dispersion along the X and Y direction for the bulk of stars around the origin of the VPD. There are 47 candidate-field stars with proper motions larger than 0.03 pixel/yr. This sample likely includes also cluster stars with large proper motions indeed, from the Galactic model by Girardi et al.\,(2005) we expect less than 20 field stars in a 2.7$\times$2.7 arcmin region in the direction of M\,2. 

The CMD of Fig.~\ref{fig:NGC7089cmd1003} reveals that M\,2 hosts a complex multiple-sequence pattern, with  a multiplicity of RGBs and SGBs and a multimodal horizontal branch (HB). The most surprising feature is a poorly-populated sequence, which runs on the red side of the majority of the stars in M\,2 and can be followed continuously from the lower part of the MS to the SGB, up to the RGB tip.
The wide separation from the bulk of M\,2 stars is an unusual feature, as multiple MSs and RGBs in `normal' GCs usually merge around the SGB. Furthermore, in GCs, the color separation between multiple MSs  typically increases when moving from the MS turn off towards fainter magnitudes, in contrast with what is observed for the tiny reddest MS in M\,2.

 An additional, sparsely-populated, SGB is clearly visible, between the main SGB and the faintest SGB. The three SGBs are  more clearly seen in the upper-right panel of Fig.~\ref{fig:NGC7089cmd1003}, which is a zoom of the left-panel
CMD around the SGB.

In order to explore the origin of the multiple sequences observed on the $m_{\rm F275W}$
 vs.\,$m_{\rm F275W}-m_{\rm F814W}$ CMD we took
advantage of the high-resolution spectroscopic data recently published by Y14. This study shows that red giants in M\,2 exhibit a multimodal abundance distribution for iron and for those neutron-capture elements that are associated with $s$-processes in solar system material.
 Specifically there is a large metallicity variation, with three groups of metal poor ([Fe/H]=$\sim -$1.7), metal-intermediate ([Fe/H]=$\sim -$1.5), and metal-rich ([Fe/H]=$\sim -$1.0) stars. Metal-intermediate stars are also enhanced in Yttrium and Zirconium by $\sim$0.5 dex with respect to the remaining stars of M\,2.

The lower-right panel of Fig.~\ref{fig:NGC7089cmd1003} shows [Y/Fe] vs.\,[Fe/H] from Y14, with black, red, and aqua symbols representing their metal-poor, metal-intermediate and metal-rich stars, respectively.
The ten stars for which {\it HST\/} F275W and F814W photometry is available are superimposed on the $m_{\rm F275W}$ vs.\,$m_{\rm F275W}-m_{\rm F814W}$ CMD. The three metallicity groups populate different RGBs: the bright, the middle, and the faint RGB correspond to metal-poor, metal-intermediate, and metal-rich  populations of Y14, respectively.

\section{Multiple populations along the RGB}
\label{sec:RGB}
A number of recent studies (see Paper\,I and references therein) have shown that the RGB, SGB, and the MS of GCs can often be separated into distinct sequences of stars, and that appropriate combinations of F275W, F336W, F438W, and F814W magnitudes are powerful tools for identifying these multiple populations.
  
The efficiency of these filters in separating different stellar populations
is closely connected to the chemical properties of GC sub-populations.
The fact that the F275W filter includes the OH molecular band, F336W the NH
band, and F438W the CH and CN bands make them very sensitive to the
effect of molecular bands in the stellar atmosphere, hence on the degree of CNO processing of the various sub-populations.   
First-generation stars are enhanced in carbon and oxygen, have low
nitrogen content, and are relatively faint in F275W and F438W, and bright in
F336W. Conversely, second-generation stars, which are carbon/oxygen poor and
nitrogen rich, are relatively bright in F275W and F438W and faint in
F336W.   
The result is that first-generation stars have bluer $m_{\rm F336W}-m_{\rm F438W}$ colors than second-generation stars at the same luminosity (Marino et al.\,2008; Bellini et al.\,2010; Sbordone et al.\,2011; Milone et al.\,2012b), while, for the same stars, the $m_{\rm F275W}-m_{\rm F336W}$ color order is reversed.
As a consequence, the $C_{\rm F275W,F336W,F438W}$=($m_{\rm F275W}-m_{\rm F336W}$)$-$($m_{\rm  F336W}-m_{\rm F438W}$) pseudo-color defined by Milone et al.\,(2013) is an extremely powerful and valuable tool to maximize the separation among the various sub-populations.    

\begin{centering}
\begin{figure*}
 \includegraphics[width=14.5cm]{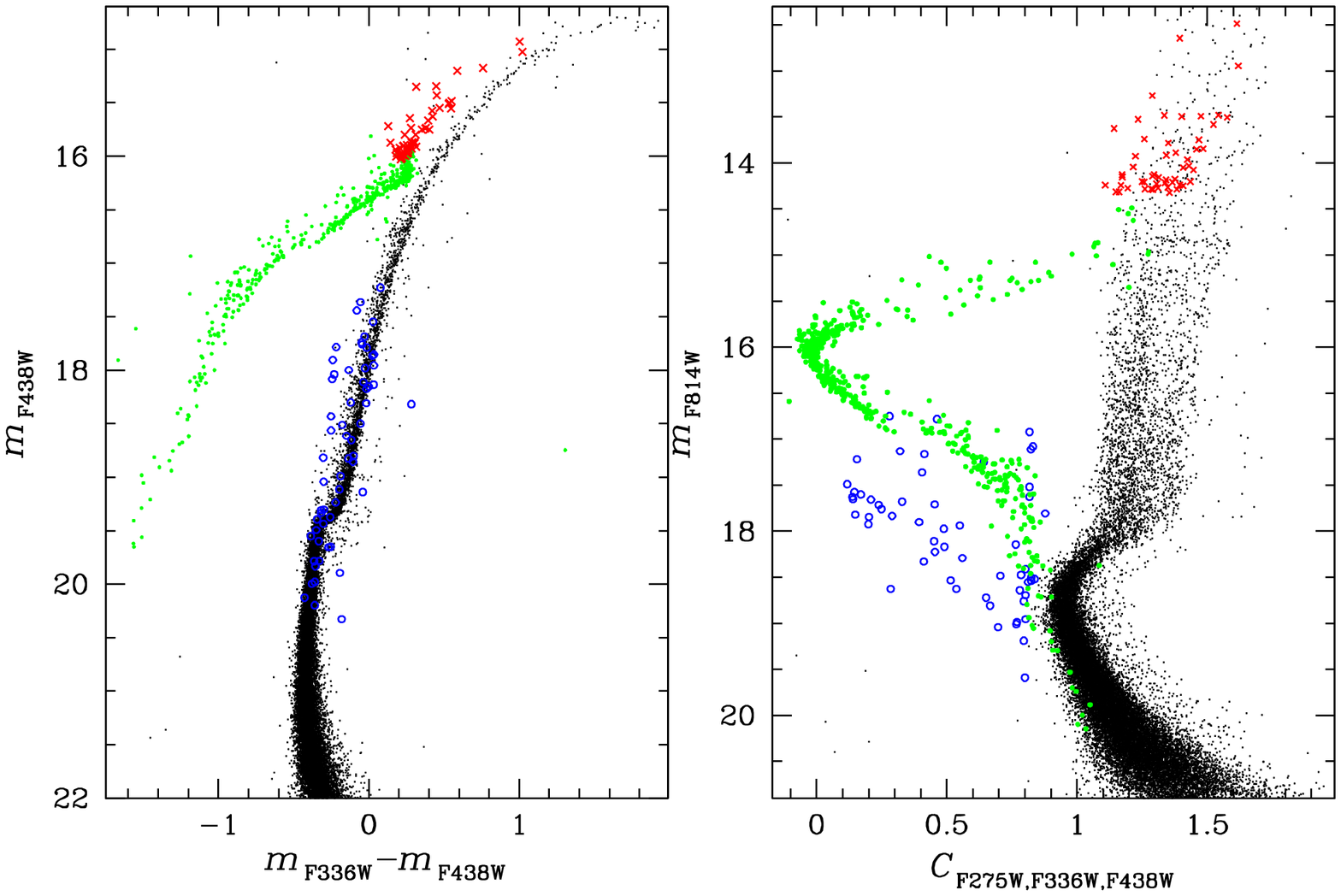}
 \caption{\textit{Left}: $m_{\rm F438W}$ vs. $m_{\rm F336W}-m_{\rm F438W}$ CMD (left panel); \textit{Right}: $m_{\rm F814W}$ vs. $C_{\rm F275W,F336W,F438W}$ diagram for M\,2 stars. AGB, HB, and BSSs, selected by eye, have been marked with red crosses, green dots, and blue circles, respectively.}
 \label{fig:NGC7089cmds}
\end{figure*}
\end{centering}

Figure~\ref{fig:NGC7089cmds} shows the $m_{\rm F438W}$ vs. $m_{\rm F336W}-m_{\rm F438W}$ CMD and the $m_{\rm F814W}$ vs. $C_{\rm F275W,F336W,F438W}$ diagram of proper-motion-selected M\,2 members. We have marked stars that, on the basis of their position in the left-panel CMD, are likely HB and asymptotic-giant branch 
(AGB) stars with green dots and red crosses, while blue stragglers (BSSs), have been selected from the right-panel diagram and have been represented with blue circles.
A visual inspection of the latter diagram reveals that the RGB of M\,2,
which shows at most a small color dispersion using  $m_{\rm F336W}-m_{\rm F438W}$ color,
spans a wide range in $C_{\rm F275W,F336W,F438W}$, with distinct RGBs.
AGB stars are also distributed
over a wide interval of $C_{\rm F275W,F336W,F438W}$, in close analogy with what is observed for the RGB, 
suggesting that the AGB of M\,2 also hosts multiple stellar populations. 
%
%
 
Multiple populations are also characterized by different helium content.
Typically,
second-generation stars are enhanced in helium and are hotter than
first-generation stars at the same luminosity.
The $m_{\rm F275W}-m_{\rm F814W}$ color is quite sensitive
to the oxygen abundance of the stellar populations through the OH molecule. In addition, the wide
color baseline provided by F275W and F814W is very sensitive to the effective
temperatures of stars, thus providing a valuable tool to identify stellar populations with different helium content or metallicity (see Milone et al.\,2012b for
details). 
  
\begin{centering}
\begin{figure*}
 \includegraphics[width=14.5cm]{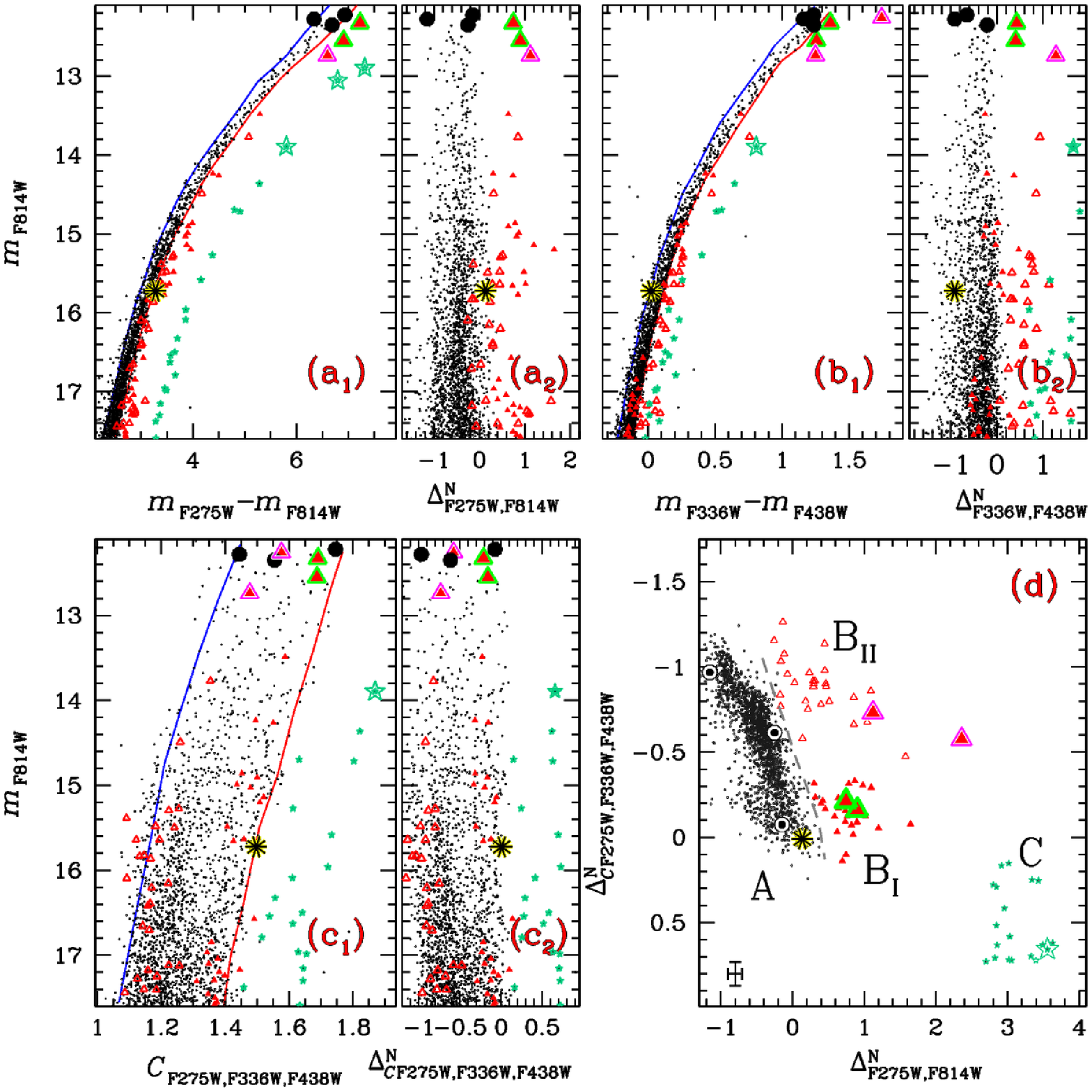}
 \caption{ Zoom-in
of the $m_{\rm F814W}$ vs. $m_{\rm
     F275W}-m_{\rm F814W}$ (panel a$_{1}$), $m_{\rm F814W}$ vs.
   $m_{\rm F336W}-m_{\rm F438W}$ CMD (panel b$_{1}$), and    vs. $C_{\rm  F275W,F336W,F438W}$ diagrams around the RGB. The two
   fiducials used to verticalize the RGB are shown as thick red and
   blue lines (see text for details). The verticalized  diagrams are plotted
   in panels a$_{2}$, b$_{2}$, and c$_{2}$. Large black dots, red triangles, and aqua stars 
        respectively     mark  the metal-poor, metal-intermediate, and metal-rich stars observed by
   Y14. Panel d shows the $\Delta_{\rm C F275W,F336W,F438W}^{\rm N}$ vs.
   $\Delta_{\rm F275W,F814W}^{\rm N}$ diagram for RGB stars.
   RGB-A, RGB-B, and RGB-C stars are colored black, red, and aqua, respectively.
 The mean error bar in shown the lower-left corner.  The magenta and green large triangles mark the  metal-intermediate/Na-rich stars and metal-intermediate/Na-poor stars, respectively.  The star analyzed by Lardo et al.\,(2012) has been represented with a large asterisk.} 
 \label{fig2}
\end{figure*}
\end{centering}

To further investigate stellar populations along the RGB of M\,2 we combined information from three distinct diagrams:
{\it (i)} $m_{\rm F814W}$ vs. $m_{\rm F275W}-m_{\rm F814W}$, 
{\it (ii)} $m_{\rm F814W}$ vs. $C_{\rm F275W,F336W,F438W}$, 
{\it (iii)} $m_{\rm F814W}$ vs. $m_{\rm F336W}-m_{\rm F438W}$. 
 We show these three CMDs in the panels a$_{1}$,  b$_{1}$, and c$_{1}$ of Fig.~\ref{fig2}, respectively. In the rest of this Section, we have excluded AGB and HB stars, and restricted our analysis to RGB stars with 12.1$<m_{\rm F814W}<$17.6.
We started by deriving by hand the blue and the red fiducial lines that are shown in panels a$_{1}$, b$_{1}$, and c$_{1}$ of Fig.~\ref{fig2} and mark the blue and the red envelope of the  main RGB. 
The fiducials are then used to rectify the RGB in such a way that the blue and
the red fiducials translate into vertical lines with abscissa equal to $-$1 and 0, respectively. The abscissa in panels a$_{2}$, b$_{2}$, and c$_{2}$ of Fig.~\ref{fig2} are named $\Delta_{\rm  F275W,F814W}^{\rm N}$, $\Delta_{\rm F336W,F438W}^{\rm N}$, and $\Delta_{\rm C  F275W,F336W,F438W}^{\rm N}$, respectively, 
 and have been calculated for each star as: 
$\Delta_{\rm X}^{\rm N}$=[($X-X_{\rm blue~fiducial}$)/($X_{\rm red~fiducial}-X_{\rm blue~fiducial}$)]-1 where $X$=($m_{\rm F275W}-m_{\rm F814W}$), ($m_{\rm
     F336W}-m_{\rm F438W}$), or $C_{\rm  F275W,F336W,F438W}$.

Panel d of Fig.~\ref{fig2} shows $\Delta_{\rm C F275W,F336W,F438W}^{\rm N}$ as a function of $\Delta_{\rm F275W,F814W}^{\rm N}$. 
 In this section we exploit this diagram to identify the three main stellar components of M\,2 (A, B, and C) and the two sub-populations of the component B (namely $B_{\rm I}$, and $B_{\rm II}$) along the RGB.
Most of the stars, including those of the most metal-poor metallicity
peak detected by Y14, are distributed in the left region of the diagram 
 and follow a well-defined pattern. In this paper we will  refer to these stars as population A and represent them with black symbols. Noticeably, a small fraction of RGB stars in M\,2 exhibit large values of $\Delta_{\rm F275W,F814W}^{\rm N}$ and $\Delta_{\rm C F275W,F336W,F438W}^{\rm N}$ and are separated from RGB-A stars by the dashed line that we have drawn by hand. We have arbitrarily divided stars on the right of this line into two samples: {\it i)} a group of stars with $\Delta_{\rm F275W,F814W}^{\rm N}>2.5$, which we have represented with aqua symbols and  named population C, and {\it ii)} another group with intermediate $\Delta_{\rm F275W,F814W}^{\rm N}$ values, marked with red triangles and  named population B. 
These colors and symbols are used consistently in the other panels of this figure. 

Because of their colors, stars on  RGB-C correspond to the reddest RGB discussed in Sect.~\ref{sec:CMD} and include the only metal-rich spectroscopic 
targets for which F275W, F336W, F438W, and F814W photometry is available. 

 RGB-B has bluer $m_{\rm F275W}-m_{\rm F814W}$ and $m_{\rm F336W}-m_{\rm F438W}$ colors with respect  than the RGB-C, but it is redder than the RGB-A.
RGB-B stars span a wide $m_{\rm F275W}-m_{\rm F814W}$ and $m_{\rm F336W}-m_{\rm F438W}$ color range and  appear to cluster around two different regions in the $\Delta_{\rm C F275W,F336W,F438W}^{\rm N}$ vs.\,$\Delta_{\rm F275W,F814W}^{\rm N}$ plane, 
 indicating the existence of at least two stellar populations  within
group B. We name them populations B$_{\rm I}$ and B$_{\rm II}$
as illustrated in Fig.~\ref{fig2}. 
The four metal-intermediate targets studied by Y14 for which photometry is available belong to the RGB-B group. Among them, two stars are highly enhanced in sodium ([Na/Fe]$\sim$0.55) and the other two have lower sodium abundance ([Na/Fe]$\sim$0.15). Na-rich and Na-poor stars, marked as magenta and large green triangles the panel (d) of Fig.~\ref{fig2} occupy the B$_{\rm I}$ and B$_{\rm II}$ regions in the $\Delta_{\rm C F275W,F336W,F438W}^{\rm N}$ vs. $\Delta_{\rm F275W,F814W}^{\rm N}$  diagram, respectively.
This fact confirms that these star groups represent different stellar sub-populations that have different sodium abundance.
 
 Lardo et al.\,(2012) have performed a CN- and CH-index study of 38 red giants of M\,2 and have identified two groups of CN-strong (CH-weak) and CN-weak (CH-strong) stars.  Furthermore, they have determined the abundance of carbon and nitrogen in 35 stars and found significant star-to-star variations of both elements at all the luminosities that form an extended C-N anticorrelation.
 {\it HST} photometry is available for only one star (\#6609) analyzed by Lardo and collaborators. It is CN weak (Lardo et al.\,2012), belongs to population A, and has been represented with a black large asterisk in Fig.~\ref{fig2}. Unfortunately, Lardo et al.\,(2012) have not inferred C and N abundances for this star because of the low signal-to-noise ratio of their spectra.

From the number of stars in the different groups of 
panel d of Fig.~\ref{fig2} we infer that in the central region of M\,2
96.1$\pm$2.2\% of stars belong to population A, 2.9$\pm$0.4\% to population B, and 1.0$\pm$0.2\% to population C.
Populations B$_{\rm I}$ and B$_{\rm II}$, contain approximately the same number of stars (48$\pm$12\% and 52$\pm$12\% of RGB-B stars, respectively).

\begin{centering}
\begin{figure*}
 \includegraphics[width=14.5cm]{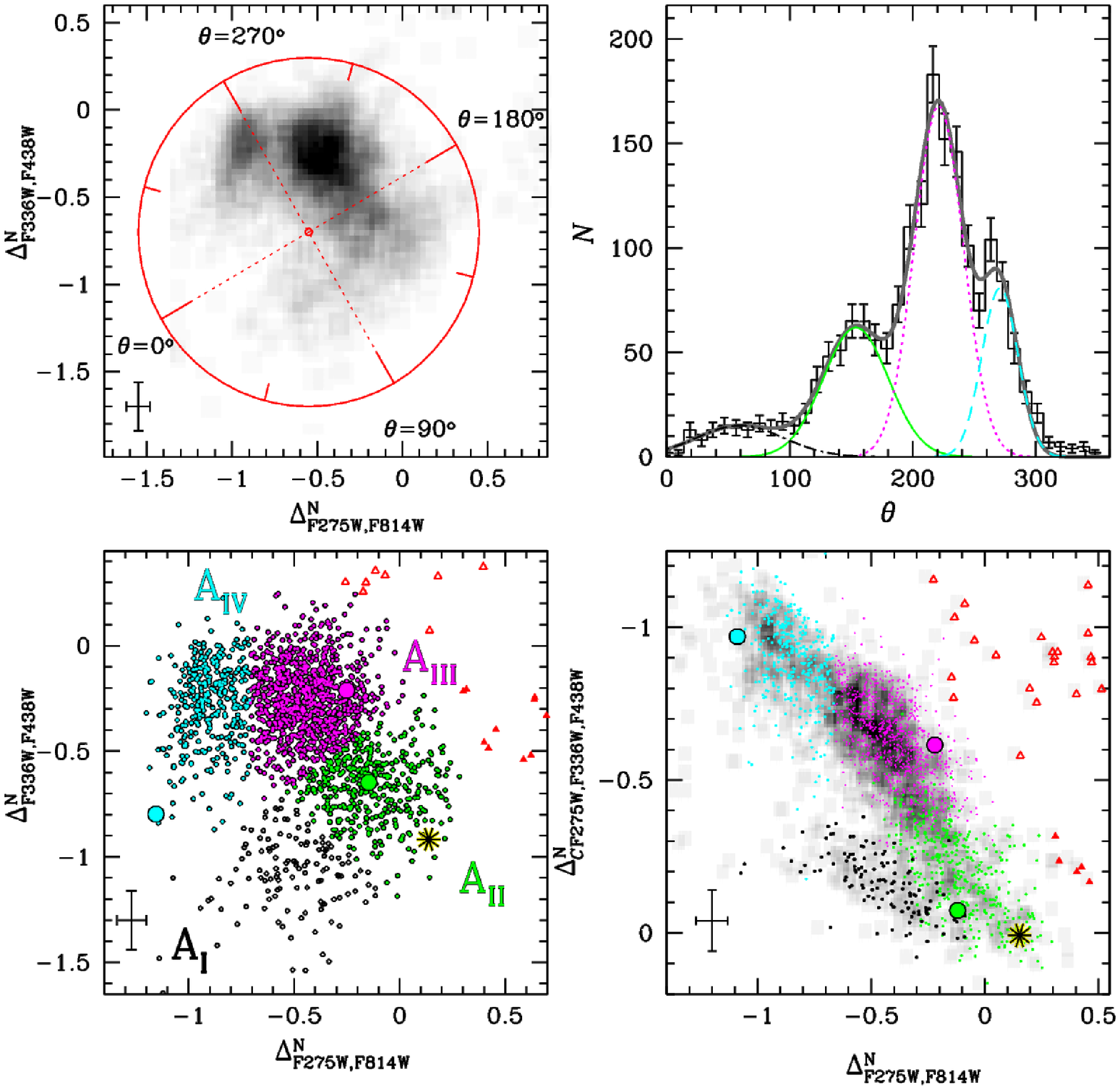}
 \caption{\textit{Upper-left panel}: $\Delta_{\rm F336W,F438W}^{\rm N}$ vs. $\Delta_{\rm  F275W,F814W}^{\rm N}$ Hess diagram for stars in the RGB-A. The reference frame adopted to measure the  polar angle, $\theta$, for RGB-A stars is also represented. The long- and short-notched lines mark different angles, from $\theta$=0$^{\rm o}$ to $\theta$=360$^{\rm o}$ in steps of 90$^{\rm o}$, and 45$^{\rm o}$, respectively. See text for details. 
\textit{Upper-right
panel}: Histogram distribution of $\theta$ for RGB-A stars. The least-squares best-fit multi-Gaussian function is represented with gray line, while its four components are colored black, green, magenta, and cyan.
\textit{Lower panels}: $\Delta_{\rm F336W,F438W}^{\rm N}$ vs.\,$\Delta_{\rm  F275W,F814W}^{\rm N}$ diagram for RGB-A stars (left panel). Stars observed by Y14 are represented with large symbols. We have defined four groups of stars (A$_{\rm I}$-A$_{\rm IV}$), and colored them black, green, magenta, and cyan, respectively. The same stars are plotted
with the same color codes in the  $\Delta_{\rm C F275W,F336W,F438W}^{\rm N}$ vs. $\Delta_{\rm F275W,F814W}^{\rm N}$ diagram shown in the lower-right panel. The filled and open red triangles represent RGB-BI and RGB-BII stars, respectively.  The mean error bar is plotted in the lower-left corner of each panel.}
 \label{f3}
\end{figure*}
\end{centering}
%
 The upper-left panel of Fig.~\ref{f3} shows the Hess diagram in the
 $\Delta_{\rm F336W,F438W}^{\rm N}$ vs.\,$\Delta_{\rm  F275W,F814W}^{\rm N}$ plane.
Here RGB-A stars exhibit an even more complex pattern, defining a kind of semi-circle. 

About half of the stars are clustered around ($\Delta_{\rm F275W,F814W}^{\rm N}$,$\Delta_{\rm  F336W,F438W}^{\rm N}$) $\simeq$($-$0.5,$-$0.2). The presence of 
three additional clumps suggest that the RGB-A hosts  four stellar sub-populations  and we use the diagram plotted in the lower-left panel to define them. To do this, we have somewhat arbitrarily separated four groups of stars in  the $\Delta_{\rm F336W,F438W}^{\rm N}$ vs.\,$\Delta_{\rm  F275W,F814W}^{\rm N}$ diagram
 and we have named 
them
A$_{\rm I}$, A$_{\rm II}$, A$_{\rm III}$, and A$_{\rm IV}$, and colored them in black, green, magenta, and cyan, respectively. The same colors are used consistently in the $\Delta_{\rm C F275W,F336W,F438W}^{\rm N}$ vs.\,$\Delta_{\rm  F275W,F814W}^{\rm N}$ diagram plotted in the lower-right panel.

In order to estimate the fraction of stars in each sub-population, we have introduced the  polar reference  frame shown in the upper-left panel of Fig.~\ref{f3}.
To do this, we have arbitrarily translated the origin to 
the
point indicated by the red circle ($\Delta_{\rm F275W,F814W}^{\rm N}$,$\Delta_{\rm  F336W,F438W}^{\rm N}$)$=$($-$0.55,$-$0.7) and rotated the axes by 208$^{\rm o}$ counterclockwise.
The histogram distribution of  the polar angle, $\theta$, in the upper-right panel clearly shows four main peaks. We have fitted this distribution %
 with a least squares fit of the sum of four Gaussians and estimated the fraction of stars in each component from the  area of the four Gaussians.
We infer that 8$\pm$3\%, 23$\pm$4\%, 51$\pm$5\%, and 18$\pm$4\% of stars belong to 
 populations A$_{\rm I}$-A$_{\rm IV}$, respectively.

 The nature of these different stellar populations on RGB-A can be clarified by  again combining our photometric results, with the chemical abundances from the spectroscopy in Y14.
The green, magenta, and cyan circles  superimposed on the diagram of Fig.~\ref{f3} refer to sodium-poor stars ([Na/Fe]=$-$0.16$\pm$0.13), sodium-intermediate ([Na/Fe]=0.18$\pm$0.13), and sodium-rich ([Na/Fe]=0.35$\pm$0.13) in Y14, respectively.  The star marked by the black asterisk is CN weak  (Lardo et al.\,2012) and belongs to population A$_{\rm II}$.
The fact that populations A$_{\rm II}$ and A$_{\rm III}$ host the Na-poor and the Na-intermediate stars,  respectively, suggests
that these two stellar populations host stars with different light-element abundances. The most Na-rich star of Y14 
is not clearly associated to any bump of stars, 
 although it is possible that it 
might be 
related to population A$_{\rm IV}$,  at least for the $\Delta_{\rm C F275W,F336W,F438W}^{\rm N}$ value. 
 Its anomalous position in the $\Delta_{\rm F275W,F814W}^{\rm N}$ and $\Delta_{\rm F336W,F438W}^{\rm N}$ 
diagrams 
may be intrinsic, but we cannot exclude the possibility that it is due to photometric errors.

\section{Multiple populations along the MS}
\label{sec:MS}
In section~\ref{sec:CMD} we have shown that %
 M\,2 includes a poorly-populated sequence of stars that is associated with population C (hereafter MS-C) and is clearly visible in the CMD of Fig.~\ref{fig:NGC7089cmd1003} on the redward side of the main MS.
 We postpone the analysis of MS-C to the next section and analyze in %
 this section 
 the bulk of MS stars which are mostly associated with population A (MS-A).
 It is not possible to identify population-B stars along the MS from our dataset.

\begin{centering}
\begin{figure*}
 \includegraphics[width=14.5cm]{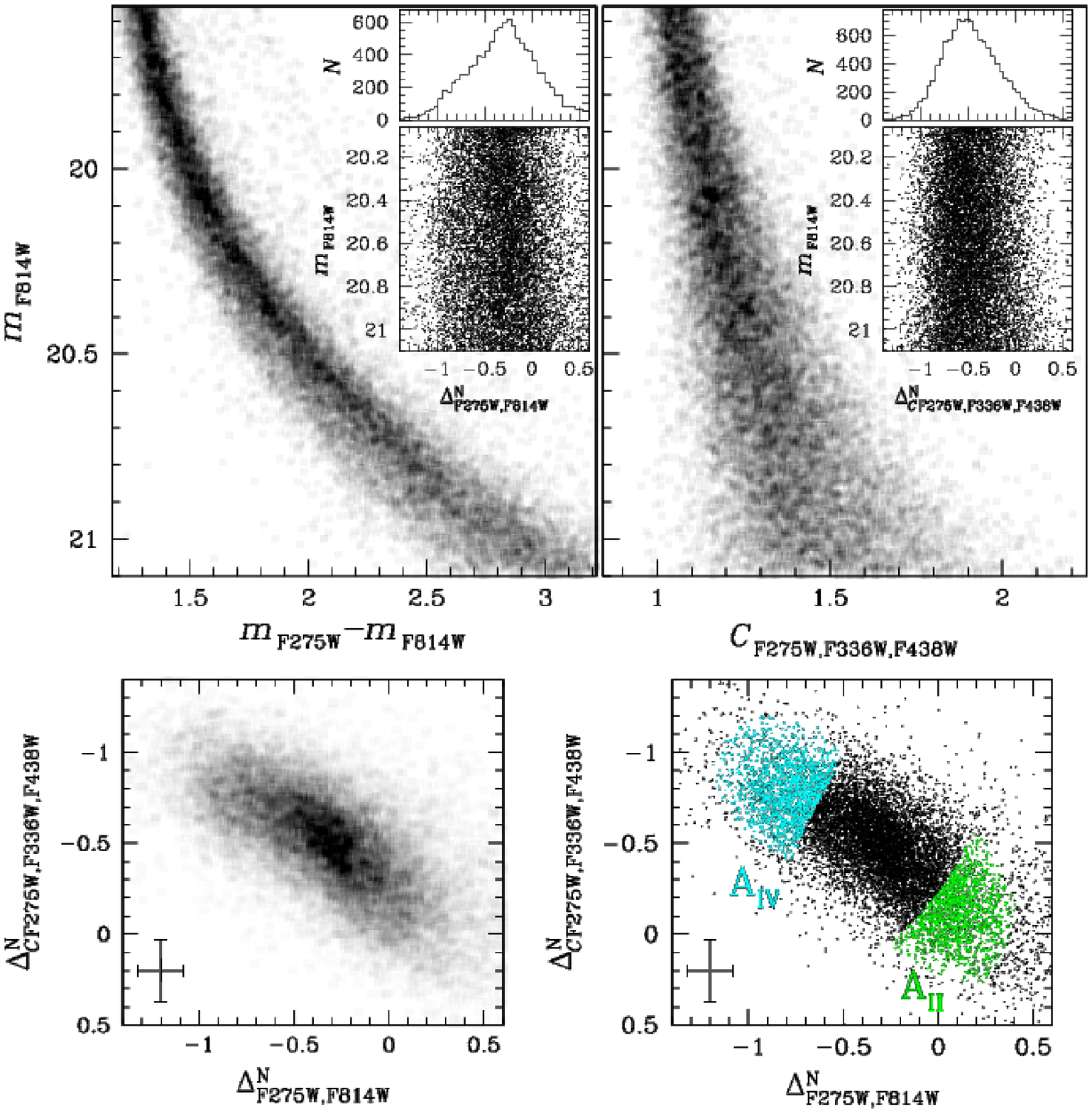}
 \caption{\textit{Upper panels}: $m_{\rm F814W}$ vs.\,$m_{\rm F275W}-m_{\rm F814W}$ (left), $m_{\rm F814W}$ vs. $C_{\rm F275W,F336W,F438W}$ (right) diagrams of MS stars. The insets show $m_{\rm F814W}$ against $\Delta^{\rm N}_{\rm F275W, F814W}$ and $m_{\rm F814W}$ against $\Delta^{\rm N}_{\rm C~F275W,F336W,F438W}$ for stars with  20.05$<m_{\rm F814W}<$21.10, and the corresponding histogram distribution. 
The $\Delta^{\rm N}_{\rm F275W, F814W}$ and  $\Delta^{\rm N}_{\rm C~F275W,F336W,F438W}$ have been obtained by subtracting to each star the color of the fiducial line drawn by hand to reproduce the CMD.
\textit{Lower-left panel}: 
 $\Delta^{\rm N}_{\rm C~F275W,F336W,F438W}$ vs. $\Delta^{\rm N}_{\rm F275W, F814W}$ 
Hess diagram.
In the right panel we have colored green and cyan the stars 
we consider to be the progeny of group A$_{\rm II}$ and A$_{\rm IV}$ identified along the RGB. 
}
 \label{fig3}
\end{figure*}
\end{centering}
In order to identify 
possible multiple
stellar populations along MS-A of M\,2, we followed
the same approach adopted in Sect.~\ref{sec:RGB} for the RGB. The $m_{\rm F814W}$ vs. $m_{\rm F275W}-m_{\rm F814W}$ Hess diagram for MS stars is shown in the left panel of Fig.~\ref{fig3}. Below $m_{\rm F814W} \sim $20, the MS of M\,2 is widely spread in color with a tail of stars on the blue side of the bulk of MS stars. The color separation between the red and the blue sides of MS-A is about 0.2 mag at $m_{\rm F814W} \sim$20 and increases towards fainter magnitudes, reaching $\sim$0.5 mag at $m_{\rm F814W} \sim$21. 
It is significantly larger than the $m_{\rm F275W}-m_{\rm F814W}$ color error, which 
 is between 0.03 and
0.08 mag in the same magnitude interval\footnote{The $m_{\rm F275W}-m_{\rm F814W}$ error has been estimated as $\delta(m_{\rm F275W}-m_{\rm F814W}) = \sqrt{\delta(m_{\rm F275W})^{2}+\delta(m_{\rm F814W})^{2}}$, where $\delta(m_{\rm F275W})$ and $\delta(m_{\rm F814W})$ are the r.\,m.\,s.\, of magnitude measurements from the six F275W and five F814W exposures, respectively divided by the square root of the  number of images minus one.}.
The verticalized $m_{\rm F814W}$ vs. $\Delta^{\rm N}_{\rm F275W, F814W}$ is plotted in the inset for stars with  20.05$<m_{\rm F814W}<$21.1.
Here,
the MS-A spread is more-clearly visible. The histogram distribution of $\Delta^{\rm N}_{\rm F275W, F814W}$ is skewed towards the blue, 
further
confirming the presence of a blue tail of stars.

The $m_{\rm F814W}$ vs.\,$C_{\rm F275W,F336W,F438W}$ Hess diagram of MS stars in the  upper-right  
panel of Fig.~\ref{fig3} furthermore reveals a red tail of stars, confirmed by the  $m_{\rm F814W}$ vs. $\Delta^{\rm N}_{\rm C~ F275W,F336W,F438W}$ verticalized diagram and the histogram of the $\Delta^{\rm N}_{\rm C~ F275W,F336W,F438W}$ distribution shown in the inset. 
 In the lower-left panel we plot $\Delta^{\rm N}_{\rm C~ F275W,F336W,F438W}$ as a function of $\Delta^{\rm N}_{\rm F275W, F814W}$ for MS-A stars with  20.05$<m_{\rm F814W}<$21.10. The comparison of this figure with the corresponding plot shown in the lower-right panel of Fig.~\ref{f3} for the RGB reveals that MS-A stars and the RGB-A share similarities. In both cases there is a tail of stars with small $\Delta^{\rm N}_{\rm F275W, F814W}$ and $\Delta^{\rm N}_{\rm C~ F275W,F336W,F438W}$ and  second one with large $\Delta^{\rm N}_{\rm F275W, F814W}$ and $\Delta^{\rm N}_{\rm c~ F275W,F336W,F438W}$.

\begin{centering}
\begin{figure*}
 \includegraphics[width=14.5cm]{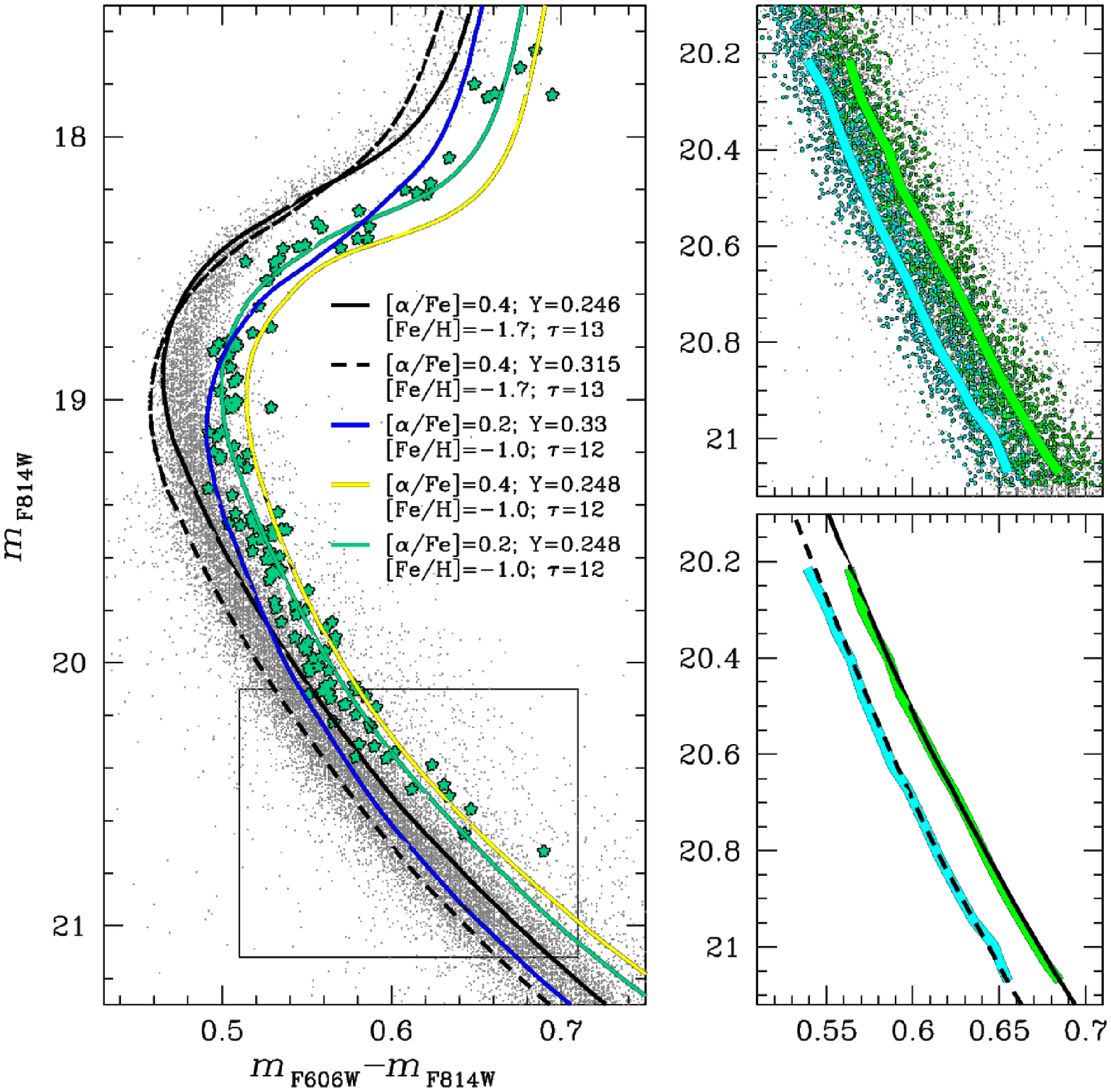}
 \caption{\textit{Left}: $m_{\rm F814W}$ vs. $m_{\rm F606W}-m_{\rm
     F814W}$ CMD of M\,2, with five superimposed isochrones with
   the values of [Fe/H], [$\alpha$/Fe], age, and Y listed in the figure inset.
 Aqua symbols highlight population C stars.
 {\it Right panels} Zoom-in of the same CMD shown in the left
panel around  the MS (upper panel). A$_{\rm II}$ and A$_{\rm IV}$ stars, as
        defined in Fig.~\ref{fig3}, are plotted with green and cyan points, respectively. The  green and the cyan line are the fiducials of A$_{\rm II}$ and A$_{\rm IV}$ MSs. In the lower-right panel we have superimposed the two metal-poor isochrones on the fiducials.  }
 \label{fig4}
\end{figure*}
\end{centering}

To further investigate multiple populations along the MS-A, we selected by eye two groups of stars with extreme $m_{\rm F275W}-m_{\rm F814W}$ and $C_{\rm F275W,F336W,F438W}$  and highlighted them in the lower-right panel of Fig.~\ref{fig3} with cyan and green colors, respectively. To minimize the effect of binaries or stars with large photometric errors we have excluded stars with very large, and very small $m_{\rm F275W}-m_{\rm F814W}$ and $C_{\rm F275W,F336W,F438W}$ values. We will use the same color code in Fig.~\ref{fig4}.

Milone et al.\,(2013) have shown that  RGB stars with both extreme $m_{\rm F275W}-m_{\rm F814W}$ and $C_{\rm F275W,F336W,F438W}$ values are the progeny of MS stars with correspondingly extreme $m_{\rm F275W}-m_{\rm F814W}$ and $C_{\rm F275W,F336W,F438W}$. %
According to this scenario, the cyan and green stars selected in Fig.~\ref{fig3} correspond to the groups of stars that have been defined as A$_{\rm IV}$ and A$_{\rm II}$, respectively, along the RGB (see lower left panel of Fig.~\ref{f3}). 
The majority of both MS-A and RGB-A stars are clustered at intermediate $\Delta^{\rm N}_{\rm F275W, F814W}$ and $\Delta^{\rm N}_{\rm C~ F275W,F336W,F438W}$ values and  
are the progenitors of 
group A$_{\rm III}$ in the RGB (Fig.~\ref{f3}). 
The present data set does not allow us to identify the progenitors of RGB-A$_{\rm I}$ along the MS.

The extreme $\Delta^{\rm N}_{\rm F275W, F814W}$ and $\Delta^{\rm N}_{\rm C~ F275W,F336W,F438W}$ values of the A$_{\rm II}$ and A$_{\rm IV}$ stellar groups
imply that they must have extreme contents of helium and light elements.
In the next section, we will estimate
 the helium spread within MS-A.  
 We remind the reader
that both the $m_{\rm F275W}-m_{\rm F814W}$ color and the $C_{\rm F275W,F336W,F438W}$ index are sensitive to light-element abundances such as OH, CN, NH, CH molecules  of different strength affect the flux in the F275W, F336W, F438W filters.
As expected, the three groups of RGB-A stars in M\,2,
which
are clearly distinguishable in Fig.~\ref{f3}, are more confused along MS-A. This is 
 both due to the photometric errors and to a temperature difference, 
as the MS stars are hotter than the RGB.
As a consequence, light-element variations have a lower influence on the photometric passbands used in our study, due to the diminished strengths of the OH, CH and CN molecular bands (see Sbordone et al.\,2011; Cassisi et al.\,2013).

\section{Age and helium content of the stellar populations}
Y14 have determined abundances for 34 elements for 16 M\,2 red
giants, thus providing an  accurate chemical 
pattern
of the multiple stellar populations of M\,2. In the following, we will 
use their measurements
to constrain age and helium-abundance differences among stars in M\,2.
 The left panel of Fig.~\ref{fig4} shows the $m_{\rm F814W}$ vs. $m_{\rm F606W}-m_{\rm F814W}$ CMD of M\,2 from ACS/WFC photometry (Anderson et al.\,2008).
MS-C stars are highlighted with aqua diamond symbols.

In order to measure the
age and helium content, we have used the
 photometry from the F606W and F814W bands,  
as it is not
significantly affected by the light-element 
abundance variations
(Sbordone et al.\,2011; Milone et al.\,2012b).
A set of isochrones from Dotter et al.\,(2007) corresponding to different chemical composition and ages has been superimposed on the CMD of Fig.~\ref{fig4}. We have used a primordial-helium (Y=0.246) isochrone (solid black), with metallicity and  $\alpha$-element content as in Y14 ([Fe/H]=$-$1.7, [$\alpha$/Fe]=0.4) to fit population $A_{\rm II}$.

  To do this, we compared the CMD with a grid of isochrones with the same composition but different age, distance modulus, and reddening. The best fit corresponds to an apparent distance modulus ($m-M$)$_{\rm V}=15.55$, age 13.0$\pm$0.75 Gyr, and E(B$-$V)=0.07,  in agreement with values from the Harris\,(1996, 2010 update) catalog. Reddening has been converted into absorption  in the ACS/WFC F606W and F814W band as in Bedin et al.\,(2005). We assumed the same values of reddening and distance modulus for the other isochrones of Fig.~\ref{fig4}.
 The best-fit age was estimated as in Dotter et al.\,(2010) by determining the isochrone that best fit the CMD in the region between the MS turn off and the SGB. The corresponding uncertainty was inferred from the intrinsic magnitude and color spread of the MS turn off and the SGB stars and we considered as 1$\sigma$-uncertainty the range of age that envelope the bulk of these stars (see Dotter et al.\,2010 for details).

 The aqua and the yellow isochrones both have age 12.0 Gyr, primordial helium, and the same iron abundance as measured by Y14 ([Fe/H]=$-$1.0) for population C, but different $\alpha$-element content. The aqua isochrone corresponds to [$\alpha$/Fe]=0.2, consistent with spectroscopic measurements by Y14, and provides the best fit to
population C. The best-fit age has been inferred as described above for population $A_{\rm II}$ and its uncertainty corresponds to 0.75 Gyr.

As a consequence, populations A and C are consistent with being coeval within $\sim$1~Gyr and with having almost primordial helium content. For completeness, we also plot a metal-rich and helium-rich isochrone  (Y=0.33, blue line). This isochrone crosses the metal-poor one at $m_{\rm F814W}\sim$18.8, and gets bluer than it towards fainter magnitudes.
The fact that population-C stars have redder $m_{\rm F606W}-m_{\rm F814W}$ colors than the bulk of M\,2 stars at the same luminosity further supports our conclusion that the population C is not consistent with being significantly helium enhanced.
 
In the upper-right panel of Fig.~\ref{fig4} green and cyan color codes mark the two stellar groups $A_{\rm II}$ and $A_{\rm IV}$ of MS stars that we identified in Fig.~\ref{fig3}. The corresponding fiducial lines have been represented with the same colors. %
It is clear that MS-$A_{\rm II}$ stars  are redder than  stars in the MS-$A_{\rm IV}$.

In the lower-right panel, we have superimposed onto the fiducial lines 
 defined in the upper-right panel the metal-poor isochrones from the left panel,
using the same distance modulus and reddening.
 Population $A_{\rm II}$ is well fitted by an isochrone with primordial helium (Y=0.246, black continuous isochrone), [Fe/H]=$-$1.7, and [$\alpha$/Fe]=0.4.
Population $A_{\rm IV}$ is reproduced by an isochrone corresponding to a
stellar population with the same age, and the same content of iron and $\alpha$-elements as $A_{\rm IV}$, but with enhanced helium (Y=0.315, black dashed line).
We obtain the same result when we adopt BaSTI isochrones (Pietrinferni et al.\,2004, 2006). 
 The age and helium content of stellar populations inferred from isochrone fitting are summarized in Table~1 together with the values of metallicity, [$\alpha$/Fe] and [Na/Fe] used in this paper, and with the fraction of stars within each sub-population.

\begin{centering}
\begin{figure*}
 \includegraphics[width=10.5cm]{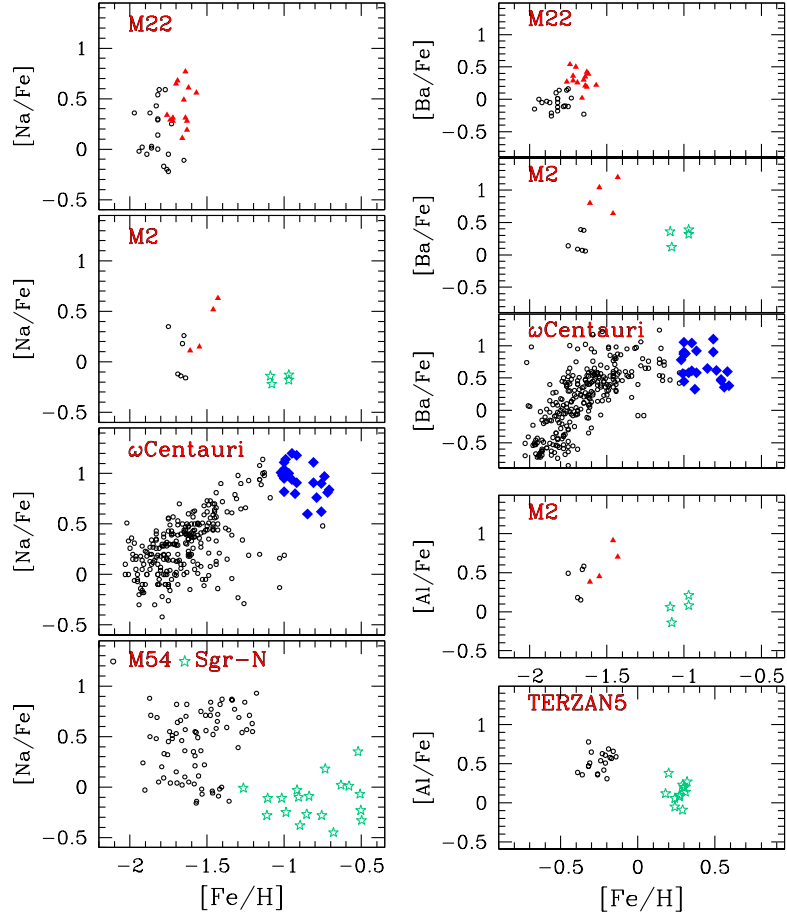}
 \caption{\textit{Left}: [Na/Fe] vs.\,[Fe/H] for four GCs with intrinsic variations in metallicity: M\,22 (Marino et al.\,2011a); M\,2 (Y14); $\omega$\,Centauri (Marino et al.\,2011b); M\,54 and the Sagittarius dwarf galaxy (Carretta et al.\,2010a). 
\textit{Right}: the three upper panels show [Ba/Fe] against [Fe/H] for M\,22 (Marino et al.\,2011a); M\,2 (Y14); $\omega$\,Centauri (Marino et al.\,2011b), while the two lower panels show [Al/Fe] vs. [Fe/H] for M\,2 (Y14) and Terzan\,5 (Origlia et al.\,2011).
The $s$-rich stars of M\,22 and M\,2 are represented with red triangles, while the blue diamonds mark stars in the RGB-a of $\omega$\,Centauri. Population-C stars of M\,2, stars in the Sagittarius dwarf galaxy nucleus, and the metal-rich stars of Terzan\,5 are plotted with aqua symbols.}
 \label{fig7}
\end{figure*}
\end{centering}

\begin{table*}
\center
\scriptsize {
\begin{tabular}{ccccccccc}
\hline
\hline
 POP & Population & [Fe/H] &  [Na/Fe] & [$\alpha$/Fe] & Y & age            &  Sect./Fig. &  Sect./Fig.\\
     &  ratio&  [dex]   &  [dex]   &  [dex]  &  [dex]   &   [Gyr]          &   RGB &  MS \\ 
\hline
$A_{\rm I}$   & 0.076$\pm$0.028   & $-$1.7 &    -    & 0.4   & -      & 13.0$\pm$0.75  & 4/4 & -- \\
$A_{\rm II}$  & 0.222$\pm$0.039   & $-$1.7 & $-$0.16 & 0.4   & 0.246  & 13.0$\pm$0.75  & 4/4 & 5/5 \\
$A_{\rm III}$ & 0.490$\pm$0.048   & $-$1.7 &    0.18 & 0.4   & -      & 13.0$\pm$0.75  & 4/4 & -- \\
$A_{\rm IV}$  & 0.173$\pm$0.036   & $-$1.7 &    0.35 & 0.4   & 0.315  & 13.0$\pm$0.75  & 4/4 & 5/5 \\
$B_{\rm I}$   & 0.014$\pm$0.003  & $-$1.5 &    0.15 & -     & - & -         & 4/3 & --  \\
$B_{\rm II}$  & 0.015$\pm$0.003  & $-$1.5 &    0.55 & -     & - & -         & 4/3 & --  \\
$C$         & 0.010$\pm$0.002    & $-$1.0 & $-$0.15 & 0.2   & 0.248  & 12.0$\pm$0.75  & 4/3 & 5/6 \\
\hline
\hline
\end{tabular}
}
\label{tab:data}
\caption{Fraction of stars, and summary of the main properties of the seven stellar populations of M\,2 used throughout this work. As discussed in the paper the values of [Fe/H], [Na/Fe], and [$\alpha$/Fe] are derived from Y14.  The last two columns indicate the section and the figure where the populations have been identified along the RGB and the MS.}
\end{table*}

\section{Comparison with other clusters}
\label{sec:discussion}
In order to better understand the properties of the seven sequences photometrically identified in M\,2, in this Section we will combine the available spectroscopic information with photometric results.
In Fig~\ref{fig7} we summarize what we currently know about clusters that show an intrinsic variation in iron content. 
Figure~\ref{fig7}
shows the  abundance of sodium, aluminum, and barium vs.\,iron content for five GCs: 
M\,2 
(Y14),
M\,22 (Marino et al.\,2011a),  M\,54 including the Sagittarius (Sgr) dwarf galaxy nucleus (Carretta et al.\,2010a), $\omega$\,Cen (Marino et al.\,2011b) and Terzan\,5 (Origlia et al.\,2011). We have represented with red
triangles stars of M\,22 and M\,2 that are enhanced in $s$-processes elements. Stars in the Sgr dwarf galaxy, in the population C of M\,2, and metal-rich stars of Terzan\,5 are plotted with aqua star symbols, while blue diamonds represent stars in the most-metal-rich RGB of $\omega$\,Cen. 

Figure~\ref{fig7} shows that populations A and B of M\,2 have different content of $s$-elements with iron-intermediate stars also being $s$-rich (Y14). 
In addition, Fig.~\ref{f3} shows that RGB-A and RGB-B stars host stellar sub-populations with different light-element abundance in close analogy with what is observed in M\,22. Y14 suggested that the populations A and B of M\,2, and other anomalous GCs (M\,22, NGC\,1851, and $\omega$\,Cen) have experienced a similar complex star-formation history.
We refer the reader to papers by  Marino et al.\,(2009, 2011a,b), Carretta et
al.\,(2010a), Da Costa \& Marino (2011), Y14 and reference therein for a
discussion of this topic. 

Y14 noticed that the RGB-C stars of M\,2 exhibit chemical properties that are rarely observed in any GC. In this section we focus on the population C of M\,2 and 
 compare it with the three extreme cases of $\omega$\,Cen, M\,54, and Terzan\,5.

\begin{itemize}
\item $\omega$\,Cen exhibits a multimodal iron distribution that spans a wide range of metallicity, with [Fe/H] ranging from $\sim -$2.0 up to $\sim -$0.7 dex (Norris \& Da Costa\,1995; Johnson \& Pilachowski\,2010; Marino et al.\,2011b; Villanova et al.\,2014).  
The MS is multimodal (Bedin et al.\,2004), 
with three main components: a red MS, which is made of
 metal-poor and helium-normal stars, a blue MS, which hosts metal-intermediate
 stars and is highly helium enhanced, by up to $Y \sim$0.39, and a metal-rich MS, named MSa in Piotto et al.\ (2005).

Bellini et al.\,(2010) have used multi-wavelength {\it HST\/} photometry to
investigate the multiple MSs in $\omega$\,Cen and showed that MSa stars have
redder $m_{\rm F275W}-m_{\rm F814W}$, $m_{\rm F336W}-m_{\rm F814W}$, and
$m_{\rm F435W}-m_{\rm F814W}$ colors than the red MS and the blue MS. They
also noticed that, when using $m_{\rm F606W}-m_{\rm F814W}$, $m_{\rm
  F625W}-m_{\rm F814W}$, or $m_{\rm F658N}-m_{\rm F814W}$ colors, MSa
becomes 
bluer than the red MS. Since MSa stars are the
progenitors of the RGBa (Pancino et al.\,2002), they are significantly more
metal-rich than the red MS. Therefore the blue colors of MSa would imply that
its stars are enriched in helium, as suggested by Norris\,(2004). 

M\,2 is similar to $\omega$\,Cen in having stellar populations that are highly
enhanced in iron relative to the others, but there are two important differences: (1) in
$\omega$\,Cen, the stars of the most metal-rich population are strongly
enhanced in sodium, aluminum and $s$-process elements (Norris \& Da
Costa\,1995; Johnson \& Pilachowski\,2010; Marino et al.\,2011b, D'Antona et
al.\,2011), in contrast with the population C of M\,2, where the content
of Na, Al, Y, and Zr are comparable with those of first-generation, 
normal-population stars (Y14, see also Fig.~\ref{fig4}). (2) The 
population C of M\,2 is not consistent with a high helium abundance. 
We conclude that $\omega$\,Cen and M\,2 have certainly experienced a different chemical-enrichment history.

\item M\,54 is another massive 
GC
with star-to-star variations in metallicity
  (Sarajedini \& Layden\,1995; Bellazzini et al.\,2008; Carretta et
  al.\,2010a,b). It lies in the nuclear region of the Sgr dwarf galaxy, although it is
  not clear whether it formed in situ or was pushed into the center by dynamical friction. 
  Carretta et al.\,(2010a,b) have derived chemical abundances for Fe, Na, and
   $\alpha$ elements for 103 red giants in the Sgr nucleus (Sgr-N). 
They used both radial velocities and photometry to identify 76 M\,54 members and 27 Sgr stars.
 Carretta and collaborators showed that M\,54 stars span a wide range in
  metallicity, with [Fe/H] ranging from $\sim -$1.9 dex up to $-$1.2 dex
  ($\sigma \sim$0.2 dex), while Sgr stars are distributed towards 
        higher iron
  content.  
In the left panels of Fig.~\ref{fig4} we plotted [Na/Fe] vs.\,[Fe/H] for RGB
  stars in M\,54 (black circles) and Sgr (aqua stars) from Carretta et
  al.\,(2010b). Stars in M\,54 exhibit a large star-to star variation in [Na/Fe], and, on average, high [Na/Fe]. By contrast, Sgr-N   stars exhibit a low sodium abundance
(see Fig.~\ref{fig7})   . 

 The stellar system including M\,54 and the Sgr-N shares some similarities with
 M\,2. M\,54 seems to include populations similar to M\,2's A and B, while the Sgr-N star chemistry is similar to the abundance pattern of M\,2 population C.
 Unlike the case of $\omega$\,Cen, however, where most of the metal-rich stars
 are strongly enhanced in sodium, both the Sgr-N stars
and  population C of M\,2 exhibit low [Na/Fe]. In addition, 
population C 
 of M\,2 is depleted in [$\alpha$/Fe] by $\sim$0.2 dex, with respect to 
  population A (Y14). This is similar to what observed in the Sagittarius
  dwarf galaxy, where the Sgr-N stars  
        have
        a lower abundance of
 $\alpha$-elements than M\,54 stars. 
 
\item 
Terzan\,5 also exhibits a very peculiar chemical composition with two
main groups of stars with [Fe/H]$\sim-$0.25 and [Fe/H]$\sim+$0.30 (Ferraro et
al.\,2009; Origlia et al.\,2011). The group of metal-rich stars has a lower
$\alpha$-element abundance with respect to the more metal-poor population,
thus suggesting that Type Ia SNe may have
played a role in the star-formation history of this cluster. 
Noticeably, at odds with the populations A and B of M\,2, there is no evidence
 for light-element variations among neither the metal-rich nor 
 the metal-poor stars of Terzan\,5. 
To further investigate similarities between M\,2 and Terzan\,5, we show in
the lower-right panels of Fig.\ref{fig7} [Al/Fe] vs. [Fe/H].  
We note that all the metal-rich stars of both M\,2 and Terzan\,5 have almost-solar
aluminum ([Al/Fe]$\sim$0.1-0.2), while the metal-poor populations of both
clusters host stars with higher aluminum content.  
A similar pattern has been observed for M\,54 and the Sgr-N with the latter
having, on average, lower aluminum content.  

\end{itemize}

\section{Summary and Conclusions}
\label{sec:discussion}
We have exploited multi-wavelength photometry from the {\it HST\/} UV Legacy Survey 
GO-13297
to investigate the stellar populations of the GC M\,2.
We have identified three main components, which we named A, B, and C. Within these three main components, we identified seven stellar sub-populations.

The main component, A, which includes 
sub-populations A$_{\rm I}$, A$_{\rm  II}$, A$_{\rm III}$, and A$_{\rm IV}$, hosts the metal-poor stars 
identified by
Y14.
 It exhibits an intrinsic spread in helium, with Y ranging from primordial
  values (Y$\sim$0.25) up to Y$\sim$0.31 and includes$\sim$96\% of stars. 
 Noticeably, the three stars with different sodium abundance identified 
by Y14 are located on the different sequences, with helium-rich stars also having  higher [Na/Fe]. 
Therefore the component A is similar to the multiple stellar populations we have identified in the majority of GCs.
        These multiple sequences host stars with the same heavy element abundance, have almost-homogeneous content
  of $s$-elements, and exhibit star-to-star variations in helium and light elements as expected for material which has been gone through hydrogen burning at high temperatures
(CNO cycle).
 
Component B has intermediate metallicity and includes $\sim$3\%of stars, is enhanced in neutron-capture elements which are usually
associated with s-processes in solar-system material,
 and includes two sub-populations, namely B$_{\rm I}$, B$_{\rm II}$,
with a different light-element abundance. It could be the analogous of the
$s$-rich and iron-rich stellar population of M\,22, and shares similarities with
the iron-rich stellar populations of $\omega$\,Cen and M\,54. 

 Component C includes $\sim$1\% of stars, is highly enhanced in iron ([Fe/H]$\sim-$1.0), and exhibits a
 lower [$\alpha$/Fe] and [Al/Fe] ratio than populations A  and B. 
Its  stars are $s$-poor and are not enhanced in helium. 
 Such properties are not compatible with self-enrichment due to either AGB or fast rotating massive stars and are not consistent with the early-disc accretion scenario.

 The combination of the photometric and spectroscopic studies of M\,2 presented in this paper  and in Y14 suggest that M\,2 is composed 
by
at least two distinct entities. The main one, containing most of the stars in M\,2, includes  
populations A and B and can be further separated into six distinct sub-populations. The minor component (population C), which makes up $\sim$1\% of the cluster stars and apparently has not produced any secondary sub-populations, is populated by stars rich in metals and with s-process elements in nearly solar proportion.
Due to the chemical properties of population-C stars, 
 one possibility is that population-C stars formed from material that made the first A sub-population but having been slightly contaminated by  supernovae  of either type.

In any event, the component A+B and the component C of M\,2 must have experienced independent star-formation histories, and as an alternative M\,2 may be final result of the merger of two stellar systems. 
The fact that M\,2 shares many similarities with the stellar system composed by M\,54 and the nucleus of the
Sagittarius dwarf galaxy, makes it very tempting to speculate that it could be the remnant of a much larger stellar system which merged with the Milky Way
in the past.

 \section*{acknowledgments}
\small
 Support for Hubble Space Telescope proposal GO-13297 was provided
  by NASA through grants from STScI, which is operated by AURA, Inc.,
  under NASA contract NAS 5-26555. APM and HJ acknowledge support by the Australian Research Council through Discovery Project grant DP120100475. %
MZ acknowledges support by Proyecto Fondecyt Regular 1110393, by the BASAL Center for Astrophysics and Associated Technologies PFB-06, and by Project IC120009 `Millennium Institute of Astrophysics (MAS)' of Iniciativa Cient{\'i}fica Milenio by the Chilean Ministry of Economy, Development and Tourism.

\bibliographystyle{aa}

\begin{thebibliography}{}
\bibitem[Anderson \& King(2003)]{2003AJ....126..772A} Anderson, J., \& King, I.~R.\ 2003, \aj, 126, 772 

\bibitem[Anderson \& King(2006)]{2006acs..rept....1A} Anderson, J., \& King, I.~R.\ 2006, Instrument Science Report ACS 2006-01, 34 pages, 1 

\bibitem[Anderson et al.(2006)]{2006A&A...454.1029A} Anderson, J., Bedin, L.~R., Piotto, G., Yadav, R.~S., \& Bellini, A.\ 2006, \aap, 454, 1029 

\bibitem[Anderson et al.(2008)]{2008AJ....135.2055A} Anderson, J., Sarajedini, A., Bedin, L.~R., et al.\ 2008, \aj, 135, 2055 

\bibitem[Anderson \& Bedin(2010)]{2010PASP..122.1035A} Anderson, J., \& Bedin, L.~R.\ 2010, \pasp, 122, 1035 

\bibitem[Bedin et al.(2004)]{2004ApJ...605L.125B} Bedin, L.~R., Piotto, G., Anderson, J., et al.\ 2004, \apjl, 605, L125 

\bibitem[Bedin et al.(2005)]{2005MNRAS.357.1038B} Bedin, L.~R., Cassisi, S., Castelli, F., et al.\ 2005, \mnras, 357, 1038 

\bibitem[Bellini \& Bedin(2009)]{2009PASP..121.1419B} Bellini, A., \& Bedin, L.~R.\ 2009, \pasp, 121, 1419 

\bibitem[Bellini et al.(2010)]{2010AJ....140..631B} Bellini, A., Bedin, L.~R., Piotto, G., et al.\ 2010, \aj, 140, 631 

\bibitem[Bellini et al.(2011)]{2011PASP..123..622B} Bellini, A., Anderson, J., \& Bedin, L.~R.\ 2011, \pasp, 123, 622 

\bibitem[Carretta et al.(2010)]{2010ApJ...714L...7C} Carretta, E., Bragaglia, A., Gratton, R.~G., et al.\ 2010, \apjl, 714, L7 

\bibitem[Carretta et al.(2010)]{2010A&A...520A..95C} Carretta, E., Bragaglia, A., Gratton, R.~G., et al.\ 2010, \aap, 520, A95 

\bibitem[Carretta et al.(2010)]{2010ApJ...722L...1C} Carretta, E., Gratton, R.~G., Lucatello, S., et al.\ 2010, \apjl, 722, L1 

\bibitem[Cassisi et al.(2013)]{2013A&A...554A..19C} Cassisi, S., Mucciarelli, A., Pietrinferni, A., Salaris, M., \& Ferguson, J.\ 2013, \aap, 554, A19 

\bibitem[Da Costa et al.(2009)]{2009ApJ...705.1481D} Da Costa, G.~S., Held, E.~V., Saviane, I., \& Gullieuszik, M.\ 2009, \apj, 705, 1481 

\bibitem[Da Costa \& Marino(2011)]{2011PASA...28...28D} Da Costa, G.~S., \& Marino, A.~F.\ 2011, PASA, 28, 28 

\bibitem[Da Costa et al.(2014)]{2014MNRAS.438.3507D} Da Costa, G.~S., Held, E.~V., \& Saviane, I.\ 2014, \mnras, 438, 3507 

\bibitem[D'Antona \& Caloi(2004)]{2004ApJ...611..871D} D'Antona, F., \& Caloi, V.\ 2004, \apj, 611, 871 

\bibitem[D'Antona et al.(2011)]{2011ApJ...736....5D} D'Antona, F., D'Ercole, A., Marino, A.~F., et al.\ 2011, \apj, 736, 5 

\bibitem[Dotter et al.(2010)]{2010ApJ...708..698D} Dotter, A., Sarajedini, A., Anderson, J., et al.\ 2010, \apj, 708, 698 

\bibitem[Ferraro et al.(2009)]{2009Natur.462..483F} Ferraro, F.~R., Dalessandro, E., Mucciarelli, A., et al.\ 2009, \nat, 462, 483 

\bibitem[Girardi et al.(2005)]{2005A&A...436..895G} Girardi, L., Groenewegen, M.~A.~T., Hatziminaoglou, E., \& da Costa, L.\ 2005, \aap, 436, 895 

\bibitem[Gratton et al.(2012)]{2012A&A...544A..12G} Gratton, R.~G., Villanova, S., Lucatello, S., et al.\ 2012, \aap, 544, AA12 

\bibitem[Grundahl(1999)]{1999ASPC..192..223G} Grundahl, F.\ 1999, Spectrophotometric Dating of Stars and Galaxies, 192, 223 

\bibitem[Harris(1996)]{1996AJ....112.1487H} Harris, W.~E.\ 1996, \aj, 112, 1487 

\bibitem[Johnson \& Pilachowski(2010)]{2010ApJ...722.1373J} Johnson, C.~I., \& Pilachowski, C.~A.\ 2010, \apj, 722, 1373 

\bibitem[King et al.(2012)]{2012AJ....144....5K} King, I.~R., Bedin, L.~R., Cassisi, S., et al.\ 2012, \aj, 144, 5 

\bibitem[Lardo et al.(2012)]{2012A&A...548A.107L} Lardo, C., Pancino, E., Mucciarelli, A., \& Milone, A.~P.\ 2012, \aap, 548, A107 

\bibitem[Lardo et al.(2013)]{2013MNRAS.433.1941L} Lardo, C., Pancino, E., Mucciarelli, A., et al.\ 2013, \mnras, 433, 1941 

\bibitem[Marino et al.(2008)]{2008A&A...490..625M} Marino, A.~F., Villanova, S., Piotto, G., et al.\ 2008, \aap, 490, 625 

\bibitem[Marino et al.(2009)]{2009A&A...505.1099M} Marino, A.~F., Milone, A.~P., Piotto, G., et al.\ 2009, \aap, 505, 1099 

\bibitem[Marino et al.(2011)]{2011ApJ...731...64M} Marino, A.~F., Milone, A.~P., Piotto, G., et al.\ 2011, \apj, 731, 64 

\bibitem[Marino et al.(2011)]{2011A&A...532A...8M} Marino, A.~F., Sneden, C., Kraft, R.~P., et al.\ 2011, \aap, 532, A8 

\bibitem[Marino et al.(2014)]{2014MNRAS.442.3044M} Marino, A.~F., Milone, A.~P., Yong, D., et al.\ 2014, \mnras, 442, 3044 

\bibitem[Milone et al.(2009)]{2009A&A...497..755M} Milone, A.~P., Bedin, L.~R., Piotto, G., \& Anderson, J.\ 2009, \aap, 497, 755 

\bibitem[Milone et al.(2012)]{2012A&A...540A..16M} Milone, A.~P., Piotto, G., Bedin, L.~R., et al.\ 2012a, \aap, 540, A16 

\bibitem[Milone et al.(2012)]{2012ApJ...744...58M} Milone, A.~P., Piotto, G., Bedin, L.~R., et al.\ 2012b, \apj, 744, 58 

\bibitem[Milone et al.(2013)]{2013ApJ...767..120M} Milone, A.~P., Marino, A.~F., Piotto, G., et al.\ 2013, \apj, 767, 120 

\bibitem[Norris \& Da Costa(1995)]{1995ApJ...447..680N} Norris, J.~E., \& Da Costa, G.~S.\ 1995, \apj, 447, 680 

\bibitem[Norris(2004)]{2004ApJ...612L..25N} Norris, J.~E.\ 2004, \apjl, 612, L25 

\bibitem[Origlia et al.(2011)]{2011ApJ...726L..20O} Origlia, L., Rich, R.~M., Ferraro, F.~R., et al.\ 2011, \apjl, 726, L20 

\bibitem[Pancino et al.(2002)]{2002ApJ...568L.101P} Pancino, E., Pasquini, L., Hill, V., Ferraro, F.~R., \& Bellazzini, M.\ 2002, \apjl, 568, L101 

\bibitem[Pietrinferni et al.(2004)]{2004ApJ...612..168P} Pietrinferni, A., Cassisi, S., Salaris, M., \& Castelli, F.\ 2004, \apj, 612, 168 

\bibitem[Pietrinferni et al.(2006)]{2006ApJ...642..797P} Pietrinferni, A., Cassisi, S., Salaris, M., \& Castelli, F.\ 2006, \apj, 642, 797 

\bibitem[Piotto et al.(2005)]{2005ApJ...621..777P} Piotto, G., Villanova, S., Bedin, L.~R., et al.\ 2005, \apj, 621, 777 

\bibitem[Piotto et al.(2012)]{2012ApJ...760...39P} Piotto, G., Milone, A.~P., Anderson, J., et al.\ 2012, \apj, 760, 39 

\bibitem[Piotto et al.(2014)]{2014arXiv1410.4564P} Piotto, G., Milone, A.~P., Bedin, L.~R., et al.\ 2014, arXiv:1410.4564 

\bibitem[Sarajedini \& Layden(1995)]{1995AJ....109.1086S} Sarajedini, A., \& Layden, A.~C.\ 1995, \aj, 109, 1086 

\bibitem[Sarajedini et al.(2007)]{2007AJ....133.1658S} Sarajedini, A., Bedin, L.~R., Chaboyer, B., et al.\ 2007, \aj, 133, 1658 

\bibitem[Sbordone et al.(2011)]{2011A&A...534A...9S} Sbordone, L., Salaris, M., Weiss, A., \& Cassisi, S.\ 2011, \aap, 534, A9 

\bibitem[Yong \& Grundahl(2008)]{2008ApJ...672L..29Y} Yong, D., \& Grundahl, F.\ 2008, \apjl, 672, L29 

\bibitem[Yong et al.(2013)]{2013MNRAS.434.3542Y} Yong, D., Mel{\'e}ndez, J., Grundahl, F., et al.\ 2013, \mnras, 434, 3542 

\bibitem[Yong et al.(2014)]{2014MNRAS.441.3396Y} Yong, D., Roederer, I.~U., 
Grundahl, F., et al.\ 2014, \mnras, 441, 3396, Y14 

\bibitem[Villanova et al.(2014)]{2014ApJ...791..107V} Villanova, S., Geisler, D., Gratton, R.~G., \& Cassisi, S.\ 2014, \apj, 791, 107 

\end{thebibliography}

\end{document}